\documentclass[onecolumn,aps,pra,reprint,superscriptaddress,longbibliography]{revtex4-1}

\usepackage{amsmath}
\usepackage{graphicx}
\usepackage[lofdepth,lotdepth, caption=false]{subfig}
\usepackage{verbatim}
\usepackage{color}
\usepackage{dcolumn}
\usepackage{bm}
\usepackage{miller}
\usepackage{color}
\usepackage{amsfonts}
\usepackage{setspace}
\DeclareUnicodeCharacter{2212}{\textminus}

\usepackage[english]{babel}
\usepackage{multirow}
\usepackage{amsmath} 
\usepackage{booktabs}
\usepackage{multirow}
\usepackage{tabularx}
\usepackage[margin=1in]{geometry}

\usepackage{natbib}
\usepackage{xr-hyper}
\usepackage{hyperref}

\begin{document}

\title{Charge density wave and superconductivity modulated by c-axis stacking in the TaSe$_{2}$ polytypes}
\author{Kusal Dharmasiri}
\affiliation{Department of Physics, University of Virginia, Charlottesville, Virginia 22904, USA}
\author{Maxim Avdeev}
\affiliation{Australian Nuclear Science and Technology Organization, Locked Bag 2001, Kirrawee DC, New South Wales 2232, Australia}
\author{Despina Louca*}
\affiliation{Department of Physics, University of Virginia, Charlottesville, Virginia 22904, USA}

\begin{abstract}
The layered transition metal dichalcogenide, TaSe$_{2}$, exhibits rich electronic phenomena across its polymorphs, 1T, 2H, and 3R, largely driven by differences in atomic coordination and c-axis stacking. In the 1T phase, octahedral coordination and AA stacking promote strong interlayer coupling and stabilize a commensurate charge density wave (CDW) with star-of-David clusters that set in at high temperatures. The 2H phase exhibits trigonal prismatic coordination with AB stacking, and hosts both incommensurate and commensurate CDW phases and weak superconductivity at very low temperatures. The 3R phase, characterized by ABC stacking and trigonal prismatic coordination, exhibits enhanced superconductivity along with CDW order, attributed to modified interlayer hybridization and reduced CDW competition. These stacking-dependent variations in interlayer coupling are critical in tuning correlated states in the dichalcogenides.  

\end{abstract}

\maketitle
\section{Introduction}
\begin{figure}[h!]
\begin{center}
\includegraphics[width=7.9cm]{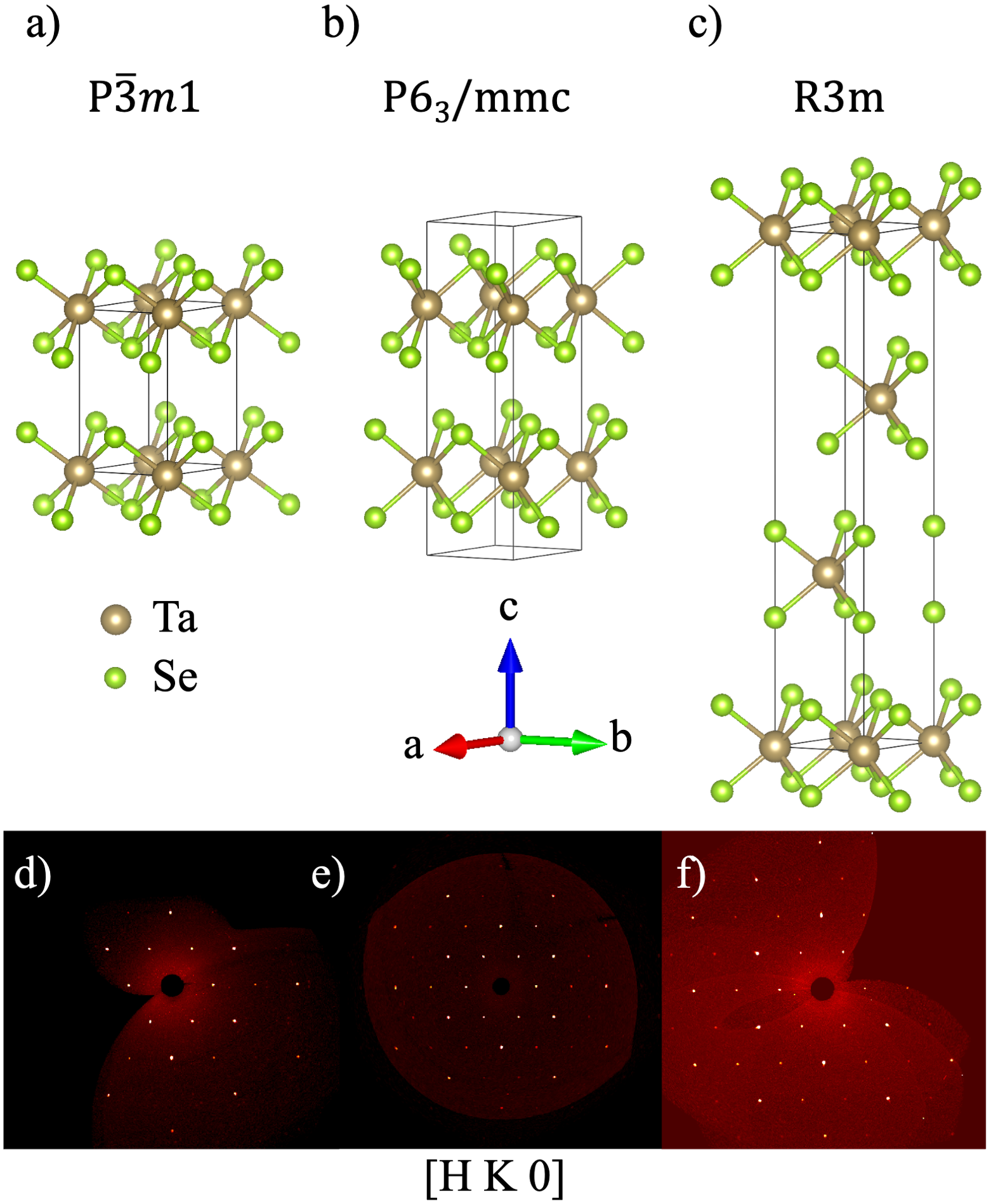}	
\end{center}
\caption{The room temperature crystal structures of the three polytypes above their CDW transition. (a) 1T - TaSe$_{2}$ with the P$\bar{3}m1$ symmetry with AA stacking; (b) The 2H - TaSe$_{2}$ with the P6$_{3}$/mmc symmetry and AB type stacking where the upper layer is rotated by 60${^0}$. (c) The 3R - TaSe$_{2}$ with ABC type stacking with each layer displaced by 1/3 along the diagonal direction resulting in the R3m symmetry. (d), (e) and (f) precision images obtained from single crystal diffraction measurement for 1T, 2H and 3R - TaSe$_{2}$ phases in the [HK0] plane at room temperature.} 
\label{fig1}
\end{figure}

Dichalcogenides have attracted considerable attention over the decades due to their unique electronic characteristics \cite{yan2015,blagojevic2024,zhang2022}. The TaSe$_{2}$ polymorphs exhibit multiple charge density wave (CDW) transitionsand various mechanism have been proposed ranging from strong electron-phonon coupling \cite{chen2018,fei2022,xu2021}, Fermi surface nesting \cite{inosov2008fermi,johannes2008fermi}, phonon softening \cite{ge2010first, shen2023precursor} and interlayer coupling and dimensionality \cite{ryu2018, ge2012effect}. In this work, we compare and contrast the three TaSe$_{2}$ polytypes where the different interlayer spacing plays an important role in correlations. In the 1T-TaSe$_{2}$ trigonal phase, each Ta atom is surrounded by six Se atoms in an octahedral geometry and the layers follow an AA type of stacking sequence (Fig.~\ref{fig1}(a)). The 1T polytype displays a commensurate CDW (CCDW) that sets in at high temperatures \cite{sayers2023exploring}. Even with the CDW transition, the system remains metallic due to partial gapping of the Fermi surface as demonstrated by angle-resolved photoemission spectroscopy (ARPES) \cite{bovet2004pseudogapped}. Insulating areas were found to coexist with metallic regions \cite{straub}. The coexistence of CDW and metallicity is unusual and indicates the presence of both localized and itinerant carriers \cite{perfetti2003,fei2022}. In its 2H polytype (hexagonal phase), TaSe$_{2}$ undergoes an incommensurate CDW (ICDW) transition around 122 K \cite{moncton1977} with an incommensurate wavevector close to $\frac{1}{3}$, followed by a commensurate phase at 90 K where the wavevector becomes $\frac{1}{3}$ \cite{leininger2011}. Each Ta atom is in a trigonal prismatic coordination with Se atoms and the layers follow an ABAB stacking sequence, where each layer is shifted relative to the one below (Fig.~\ref{fig1}(b)). This phase remains metallic even in the CDW state, with superconductivity emerging at very low temperatures due to multiple bands crossing the Fermi surface \cite{li2018,yan2015}.

More recently, a 3R polytype (rhombohedral phase) has also been synthesized that exhibits distinct CDW and superconducting properties \cite{deng2020,xie2025,huang2023}. In the 3R structure, the Ta atoms are coordinated in trigonal prisms, and the layers follow an ABC stacking sequence, repeating every three layers along the c-axis (Fig.~\ref{fig1}(c)). 
This stacking variation significantly influences the electronic properties. The superconducting transition temperature (T$_{c}$) in the 3R phase can reach up to 2.4 K, which is significantly higher than that of the 2H phase (T$_{c}$$\approx$0.15 K). The enhancement is attributed to subtle changes in the electronic structure induced by the altered stacking sequence, though the exact mechanism is not clear. It has also been proposed that CDW and superconductivity coexist in this phase. Recently, it was suggested that because 3R lacks inversion symmetry, it might enable Ising pairing, a type of spin-triplet pairing protected by strong spin-orbit coupling \cite{xie2025,luo2015}. In this work, transport measurements are combined with neutron diffraction on the three polytypes to elucidate the structural characteristics as the different CDW transitions.   

TaSe$_2$ exhibits diverse electronic, optical, and mechanical properties due to its unique layered structure and strong intra-layer bonding, while exhibiting weak inter-layer van der Waals interactions, making them suitable for applications \cite{novoselov_nd,wilson1969,chhowalla2013}. The polymorphs of Fig.~\ref{fig1} demonstrate how differences in local coordination geometries influence their electronic properties, especially the CDW behavior and superconductivity. The c-axis stacking differences among the 1T, 2H, and 3R phases directly influence the interlayer coupling and electronic band structure. The AA stacking pattern in the monolayers forms an octahedral prismatic coordination in the 1T that leads to strong interlayer orbital overlap, especially between Ta 5d orbitals \cite{ruggeri2024}, which enhances electron correlations and supports Mott-like behavior in the CDW phase. By contrast, the ABAB stacking pattern of the 2H and the ABC stacking in the 3R monolayers lead to trigonal-prismatic coordination \cite{brown1965,wilson1975}. The ABAB stacking reduces direct orbital overlap between layers, allowing for metallic conductivity even in the CDW state and enabling weak superconductivity \cite{dordevic2003, wu2019, li2023}. Meanwhile, the ABC stacking of the 3R polymorph facilitates a three-layer periodicity, which modifies the interlayer hybridization and can enhance superconductivity while supporting CDW order \cite{xie2025, li2023, huang2023, deng2020}.
Previous studies on TaSe$_2$ primarily focused on enhancing superconductivity by incorporating various substitutions or applying pressure. It was reported that incorporating sulfur (S) into the 1T - TaS$_{2-x}$Se$_{x}$, can induce superconductivity and maximum superconducting temperature can be obtained when S approaches x = 1 \cite{liu202x}. When pressure is applied on 1T-TaSe$_{2}$, the CCDW - ICDW transition temperature shifts down to $\approx$ 260 K around 4.5 GPa. As the pressure increases, the CDW phase is suppressed by 6.5 GPa, and T$_{c}$ continues to increase with pressure, reaching a maximum of 5.3 K at 15 GPa \cite{wang2017}. Furthermore, in the 2H polymorph of TaSe$_{2-x}$S$_{x}$, doping can increase superconductivity to 4.2 K when 0.52 $\leq$ x $\leq 1.65$ \cite{li2017}. In the 3R phase, substitution of Te for Se leads to an increase in T$_{c}$ as well \cite{luo2015}.
In this work, we explore the coexistence and competition between CDW order and superconductivity comparing the 3 polytypes of TaSe$_{2}$, focusing on the distinct characteristics of the single layer and of their different stacking modalities to elucidate the underlying structural effects that lead to their distinct electronic properties.

\begin{figure}[h]
\begin{center}
\includegraphics[width=7.8cm]{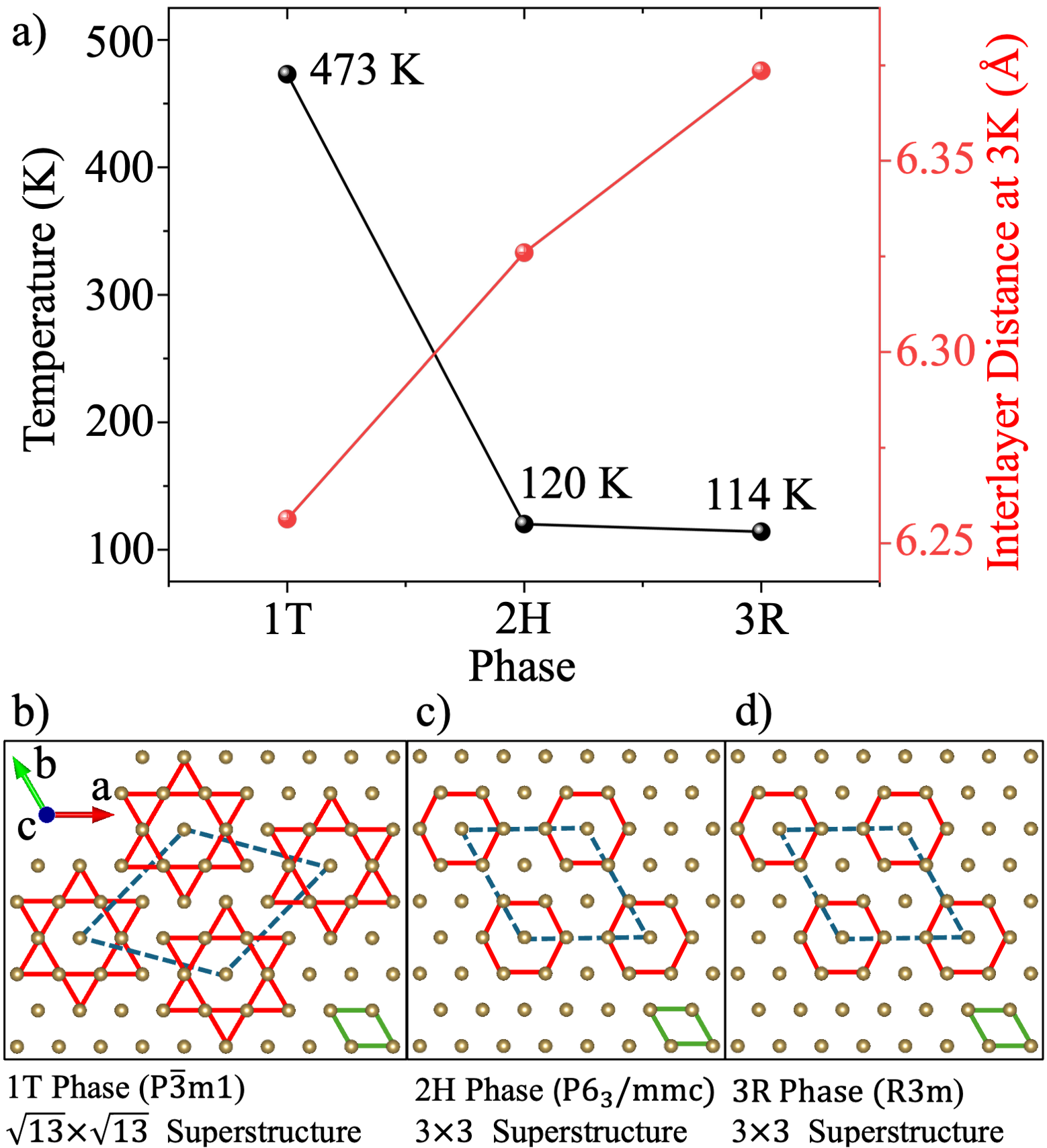}	
\end{center}
\caption{(a) A plot of the CDW transition temperatures for the 1T, 2H and 3R phases. The data for the 1T phase was obtained from Ref. \cite{samnakay2014}. The right axis shows the interlayer distance for each phase at 3 K obtained from the neutron data refinement. (b) The $\sqrt{13} \times \sqrt{13}$ CDW modulation for 1T phase. The CCDW "Star-of-David" pattern is shown in red. (c) and d) $3 \times 3$ CDW modulation for the 2H and 3R phases, respectively. The unit cells above and below the CCDW transitions are shown
with green solid and blue dashed lines, respectively.}
\label{fig2}
\end{figure}

\section{Experiment}
The TaSe${_2}$ single crystals were grown using chemical vapor transport (CVT) with iodine as the transport agent. Stoichiometric ratios of Ta and Se powders were mixed and ground, then sealed into an evacuated quartz ampoule. The ampoule was heated between 850 and 1000 K for 48 hours and quenched. The obtained powder sample was mixed with iodine (4 mgcm$^{-3}$) and placed in an evacuated quartz tube. Samples were placed in a two-zone furnace for two weeks, during which the source temperature was maintained at 1298 K for the 1T phase, 1148 K for the 2H phase and 1273 K for the 3R phase, while the growth temperature was maintained at 1223 K for the 1T phase, 1073 K for the 2H phase and 1223 K for the 3R phase. After quenching, shiny golden crystals were obtained for the 1T phase, while grayish crystals were obtained with the 2H and 3R phases. 
The phase and purity were verified using powder X-ray diffraction at room temperature. Neutron powder diffraction measurements were performed at Echidna, a high resolution powder diffractometer at the Australian Nuclear Science and Technology Organization (ANSTO). Approximately 1 gram of sample was utilized and data were collected at 3, 100, 200, and 300 K using a monochromator with a wavelength of 2.0468 \AA. The structural refinement of the neutron diffraction data was carried out using the GSAS - II package \cite{toby2013}. The results are summarized in Table I. In all the samples, the majority phase is the expected polymorph.

\begin{figure}[h!]
\begin{center}
\includegraphics[width=7.9 cm]{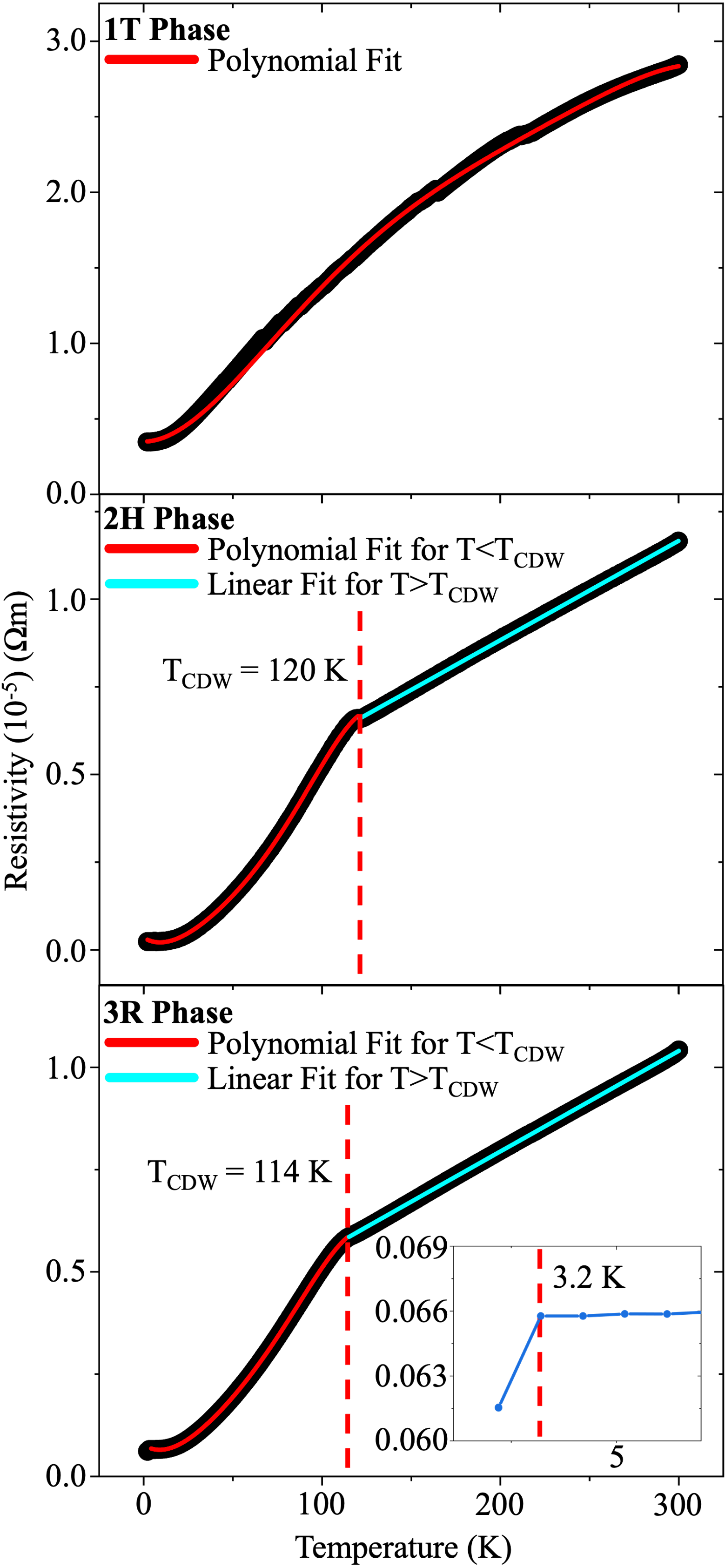}	
\end{center}
\caption{Plots of the temperature dependence of the resistivity for (a) 1T - TaSe$_2$ (b) 2H - TaSe$_2$ and (c) 3R - TaSe$_2$ single crystals. The inset shows the resistivity for 3R near the superconducting transition temperature. The fitting results for the different regions are shown in Table II.} 
\label{fig3}
\end{figure}
%

\begin{table}[t]
\centering
\caption{Analyzed values for 1T, 2H, and 3R - TaSe$_2$ samples as obtained from neutron diffraction data at 3K.}
\setlength{\tabcolsep}{4.5pt}
\renewcommand{\arraystretch}{1.2}
\begin{tabular}{|c c|c|c|c|}
\hline
\multicolumn{2}{|c|}{Sample} 
& 1T & 2H & 3R \\
\hline
\multirow{5}{*}{Phase Fraction (\%)} 
 & 1T  & 98.8 & -- & -- \\
 & 2H  & 0.5  & 100 & 5.0 \\
 & 3R  & --   & --  & 90.5 \\
 & 4Hb & 0.5  & --  & 4.0 \\
 & Se  & 0.2  & --  & 0.5 \\
\hline
\multicolumn{2}{|c|}{$R_{w}$} 
& 8.626 & 5.924 & 8.179 \\
\hline
\multicolumn{2}{|c|}{Reduced $\chi^2$}   
& 7.725    & 3.72     & 9.586 \\ 
\hline
\end{tabular}
\end{table}

\begin{table*}
\label{tab:wide_table}
\centering
\caption{Polynomial and linear fitting expressions for the different TaSe$_2$ phases.}
\renewcommand{\arraystretch}{}
\begin{tabular}{m{4cm} m{12cm}}
\hline
\multicolumn{2}{c}{\textbf{1T Phase}} \\

\textbf{Polynomial:} & \textbf{$\rho$(T)} = $0.35 - (1\times 10^{-5})\textbf{T} + (2.27\times 10^{-4})\textbf{T}^{2} - (1.72\times 10^{-6})\textbf{T}^{3} + (5.37\times 10^{-9})\textbf{T}^{4} - (6.14\times 10^{-12})\textbf{T}^{5}$ \\[1em]
\hline
\multicolumn{2}{c}{\textbf{2H Phase}} \\

\textbf{Linear:} & \textbf{$\rho$(T)} = $0.32 + (2.83\times 10^{-3})\textbf{T}$ \\[0.5em]
\hline
\textbf{Polynomial:} & \textbf{$\rho$(T)} = $0.034 - (2.98\times 10^{-3})\textbf{T} + (2.02\times 10^{-4})\textbf{T}^{2} - (3.02\times 10^{-6})\textbf{T}^{3} + (2.78\times 10^{-8})\textbf{T}^{4} - (9.96\times 10^{-11})\textbf{T}^{5}$ \\[1em]
\hline
\multicolumn{2}{c}{\textbf{3R Phase}} \\

\textbf{Linear:} & \textbf{$\rho$(T)} = $0.30 + (2.46\times 10^{-3})\textbf{T}$ \\[0.5em]
\hline 
\textbf{Polynomial:} & \textbf{$\rho$(T)} = $0.077 - (2.83\times 10^{-3})\textbf{T} + (1.91\times 10^{-4})\textbf{T}^{2} - (2.73\times 10^{-6})\textbf{T}^{3} + (2.39\times 10^{-8})\textbf{T}^{4} - (8.58\times 10^{-11})\textbf{T}^{5}$ \\[1em]
\hline
\end{tabular}
\end{table*}

\section{Results and Discussion}
The CDW is manifested as periodic modulations in both electron density and atomic positions, where the formation of a periodic lattice distortion (PLD) lowers the material's symmetry \cite{lee2025, gruner1994}. These distortions are stabilized by lattice vibrations and hybridization between Ta 5d and Se 4p orbitals \cite{ang2015}. Each phase of TaSe$_2$ exhibits distinct CDW transitions. Bulk 1T undergoes a transition from the normal state to an incommensurate CDW (ICDW) state at approximately 600 K. Subsequently, a CCDW transition occurs around 473 K, resulting in the formation of a $\sqrt{13} \times \sqrt{13}$ superstructure \cite{samnakay2014, philip2023, philip2024} with the star of David motifs. The 2H phase undergoes a transition to an ICDW at 122 K, followed by a transition to a CCDW phase at 90 K where a $3 \times 3$ superstructure is formed. Moreover, superconductivity appears below 0.14 K upon cooling \cite{bhoi2016}. The 3R - TaSe$_2$ also exhibits a $3 \times 3$ superstructure around 114 K and additionally shows superconductivity\cite{xie2025}. 

A plot of the CDW transition temperatures for each phase is shown in (Fig.~\ref{fig2}(a)), compared to the interlayer spacing from 1T (low) to 3R (high).  Also shown in (Fig.~\ref{fig2}(b) and Fig.~\ref{fig2}(c)) are the CDW modulations for each phase. It is evident that there is an inverse correlation to the interlayer coupling, where CDW transition temperature is highest when interlayer spacing is smallest and lowest when it is largest. Thus, dimensionality plays a key role in this system. Even in the monolayer, TaSe$_{2}$ retains CDW order but with a modified transition temperature \cite{yan2015, ryu2018}. It has been suggested taht the reduced dimensionality enhances electron correlations. Additionally, CDW is often observed in proximity to superconductivity, which emerges at very low temperatures \cite{bhoi2016, cho2018}. The co-existence or competition can be tuned via doping, pressure, or layer thickness, making TaSe$_{2}$ a compelling system for studying correlated electron phenomena in low-dimensional materials \cite{adam2023tuning,liu2016nature, deng2020, bhoi2016,wang2017,ryu2018}.  

In Fig.~\ref{fig3}, the resistivity $\rho$(T) plots for the 1T, 2H, and 3R-TaSe$_{2}$ phases are shown as a function of temperature. Linear and polynomial fittings were implemented for each $\rho$(T) curve, both above and below T$_{CDW}$, to elucidate its temperature dependence. The fitting parameters are listed in Table II. 1T-TaSe$_{2}$ exhibits metallic behavior and no discontinuity is observed in this temperature range since the CDW transition occurs at much higher temperatures \cite{wang2017}. The resistivity is not linear but rather has a power law dependence as seen from Table II. In the other two polymorphs a clear linear part in the resistivity at high temperatures prior to the CDW transition is observed. The 2H-TaSe$_{2}$ and 3R-TaSe$_{2}$ phases exhibit a linear part to the resistivity along with a clear drop in the linear behavior around 120 K, and 114 K, respectively, upon entering the CDW transition. Below the CDW transition, $\rho$(T) is fit with a polynomial function as shown in the Table. Note that the 2H phase exhibits superconductivity below 1 K which is not captured in this measurement. The inset in Fig.~\ref{fig3}(c) illustrates the resistivity around the superconducting transition temperature for 3R-TaSe$_{2}$. It gradually exhibits superconducting behavior around 3.2 K, closely aligning with the reported values \cite{xie2025}.

\begin{figure}[h!]
\begin{center}
\includegraphics[width=7.9cm]{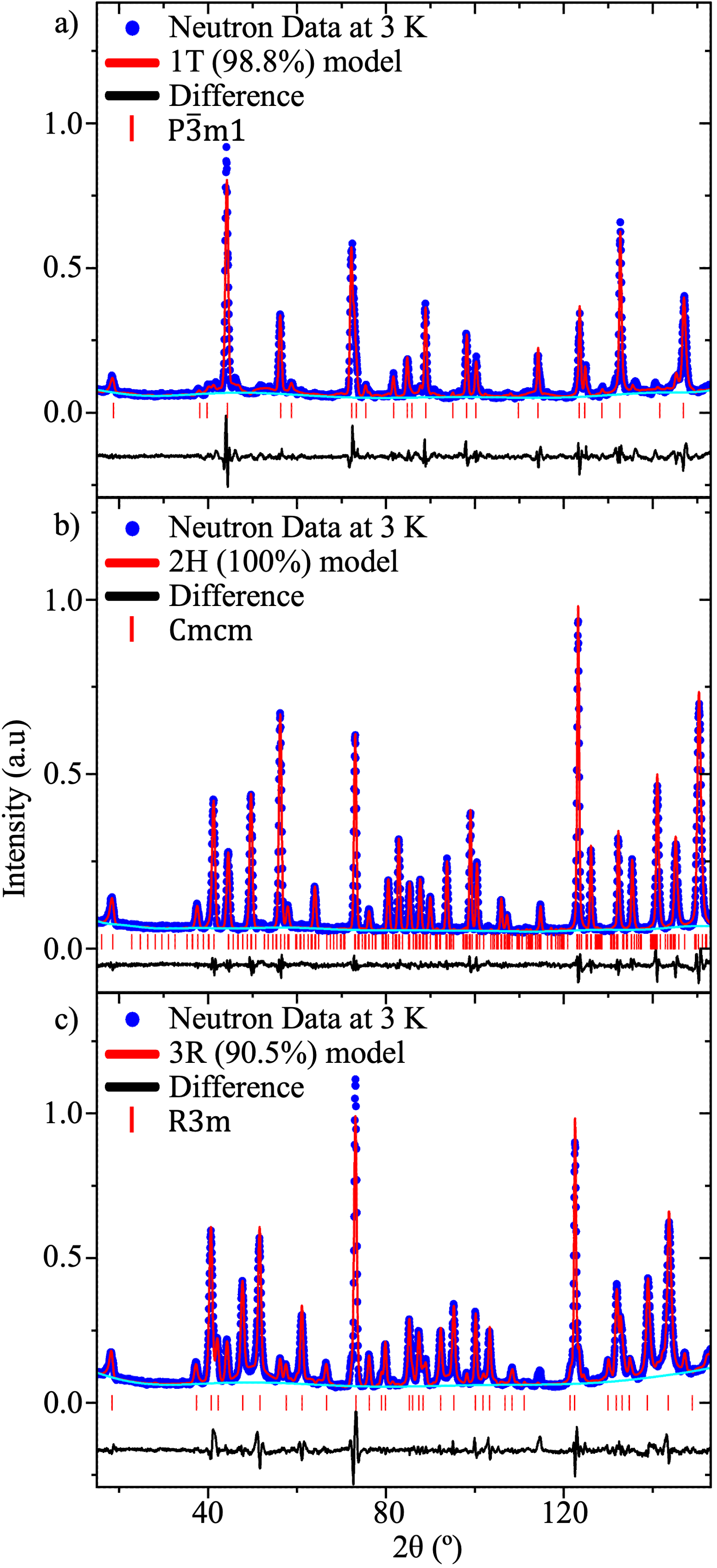}	
\end{center}
\caption{The neutron diffraction data collected at Echidna at 3 K. In (a), the neutron data for the 1T - TaSe$_{2}$ sample is compared to the model calculated using the P$\bar{3}m1$ space group. Additional phases are present in small amounts. In (b), the 2H - TaSe$_{2}$ diffraction data are compared to the model calculated using Cmcm space group. In (c), the 3R - TaSe$_2$ diffraction data is compared to the model calculated using the R3m space group. The impurities are listed in Table I.}
\label{fig4}
\end{figure}

Shown in Fig.~\ref{fig4} are plots of the diffraction intensity for the (a) 1T phase, (b) the 2H phase and (c) 3R phase from data collected at 3 K. The data were refined using the crystal symmetry shown in the figures and the results are summarized in Table I and Table III. The 2H powder data show a pure phase, while the 1T and 3R samples contain impurities, as listed in Table I. The refinement of the neutron data for the 2H phase confirms the high quality of the sample, which consists exclusively of the TaSe$_{2}$ phase with the Cmcm space group. Even though the 1T and 3R samples contain impurities, they still contain more than 90\% of the desired phase. The results from the refinemnet are summarized in Table III.

\begin{figure}[t]
\begin{center}
\includegraphics[width=7.9cm]{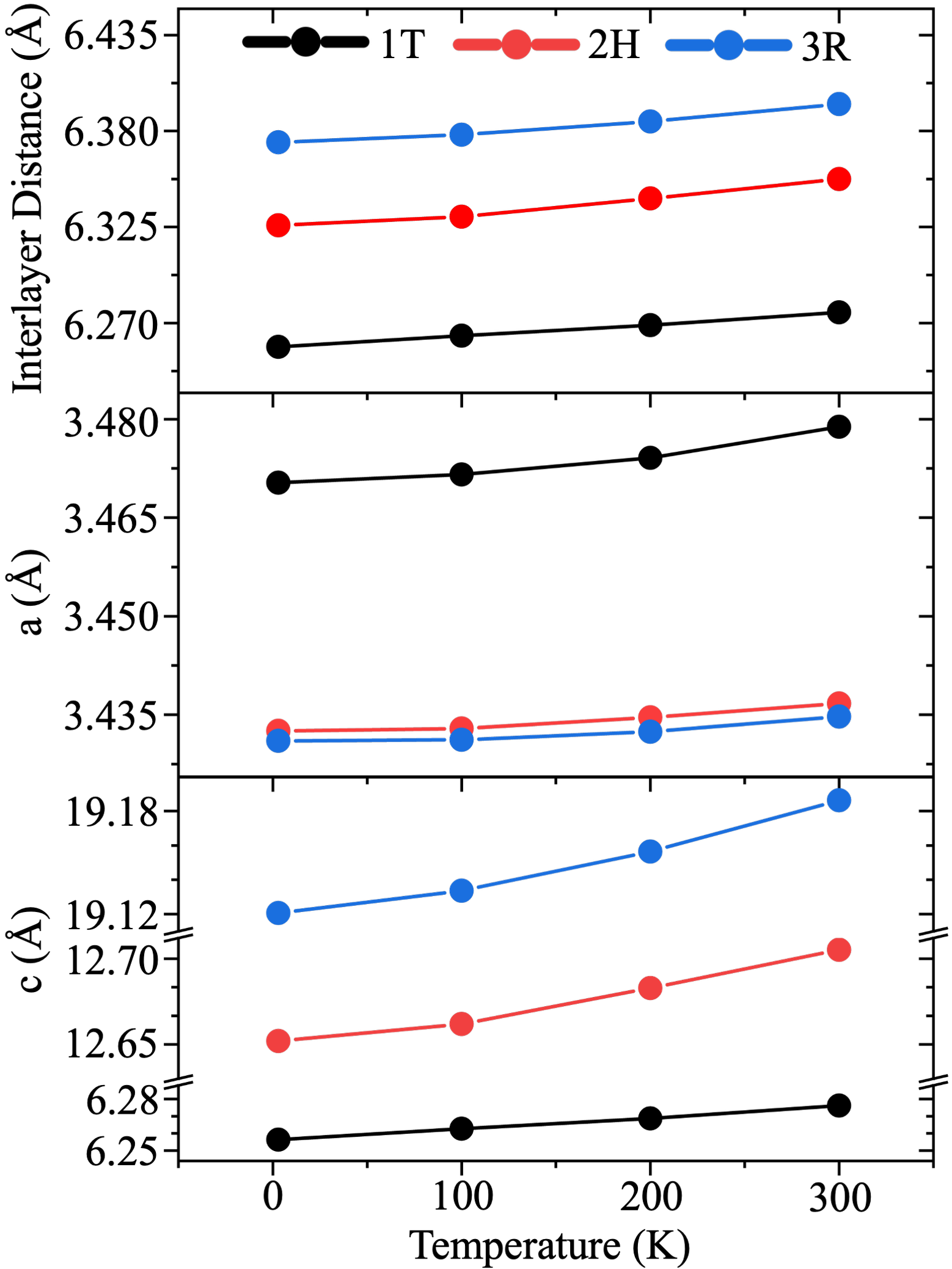}	
\end{center}
\caption{A plot of the temperature dependence of (a) the inter-layer separation, (b) the a-lattice parameter and (c) the c-lattice parameter as obtained from the refinement of the neutron diffraction data. The error bars are smaller than the size of the data points.} 
\label{fig5}
\end{figure}

The lattice parameters and interlayer spacing for the 3 polymorphs are plotted as a function of temperature and shown in Fig.~\ref{fig5}. As expected, the c-lattice constant expands proportionately to the number of layers, and is largest for the 3R phase. On the other hand, the a-lattice constant is the same for the 2H and 3R but distinctly larger for the 1T phase. Further, the interlayer spacing is smallest for the 1T, but increases with the 2H and the 3R. The increase in the interlayer spacing enhances the 2-dimensional character of the electronic structure. It is expected that the electronic correlations and orbital overlap are reduced, and can lead to quantum confinement of the carriers in the layers. This in turn can modify the band structure and bring changes to the Fermi surface topology with direct implications to superconducting pairing \cite{luo2015, luo2015superconductivity, wu2019dimensional}. In the 3R-TaSe${_2}$, the increase in interlayer spacing compared to the 2H-TaSe${_2}$ can be linked to the lower CDW transition temperature and the emergence of superconductivity with a higher T${_c}$. By contrast, 1T-TaSe${_2}$ has the smallest interlayer spacing of the three and the highest CDW transition and no superconductivity. This suggests that dimensional tuning via interlayer spacing can shift the balance between competing ground states.
There is a clear anti-correlation of the CDW transition temperature to the interlayer spacing.

The interlayer distance and the \textit{a}- and \textit{c}-lattice constants for the three polytypes, obtained from the Rietveld refinement of the neutron data from Echidna from 3 to 300 K, are shown in Figs. 4(b) and 4(c). Both the a and c lattice parameters exhibit a normal temperature dependence. Notably, 1T-TaSe$_{2}$ exhibits the largest \textit{a} lattice parameter which indicates that the Ta-Se atom arrangements in the monolayer is expanded, while the \textit{a} lattice parameters for the 2H and 3R are close indicating a similar Ta-Se atom arrangement, in support of reported data \cite{bjerkelund1967}. The differences in the single layer arrangement of the Ta and Se atoms can be seen in (new figure goes here) (need to plot each layer for the 3 phases - need to expand this). On the other hand, for the \textit{c} lattice constant, it is largest for the 3R phase, followed by the 2H and the 1T, corresponding to the number of monolayers in the unit cell. This is in agreement with previously reported values \cite{bjerkelund1967}. As the temperature increases, the separation of the monolayers is very different in the three phases (Fig. 4(a)). The monolayers are closest in the 1T phase and farthest apart in 3R.

\begin{table}[h]
\centering
\caption{Atomic coordinates and isotropic thermal factors (Uiso) for Ta and Se sites (at 3\,K).}
\label{tab:atomic_params}
\setlength{\tabcolsep}{1 pt}
\renewcommand{\arraystretch}{1.2}
\footnotesize
\normalsize
\begin{tabularx}{\linewidth}{@{}l *{3}{>{\centering\arraybackslash}X}@{}}
\hline
\textbf{1T} & \textbf{a(\AA)} & \textbf{b(\AA)} & \textbf{c(\AA)} \\
\hline
  & 3.47032(23) & 3.47032(23) & 6.25633(3)  \\
\end{tabularx}
\begin{tabularx}{\linewidth}{@{}l *{4}{>{\centering\arraybackslash}X}@{}}
\hline
& $\mathbf{x}$ & $\mathbf{y}$ & $\mathbf{z}$ & $\mathbf{U}_{\mathrm{iso}}$ \\
\hline
Ta1  & 0.00000 & 0.00000 & 0.00000 & 0.03114(53) \\
\hline
Se1  & 0.33333  & 0.66667 & 0.2645 & 0.00909(42) \\
\hline
\end{tabularx}
\begin{tabularx}{\linewidth}{@{}l *{3}{>{\centering\arraybackslash}X}@{}}
\hline
\textbf{2H} & \textbf{a(\AA)} & \textbf{b(\AA)} & \textbf{c(\AA)} \\
\hline
  & 10.297609(52) & 17.835982(90) & 12.652424(12)  \\
\end{tabularx}
\begin{tabularx}{\linewidth}{@{}l *{4}{>{\centering\arraybackslash}X}@{}}
\hline
& $\mathbf{x}$ & $\mathbf{y}$ & $\mathbf{z}$ & $\mathbf{U}_{\mathrm{iso}}$ \\
\hline
Ta1  & 0.00000 & 0.66676 & 0.25000 & 0.00839(30) \\
Ta2  & 0.00000 & 0.33324 & 0.25000 & 0.01094(32) \\
Ta3  & 0.00000 & 0.00001 & 0.25000 & 0.00815(31) \\
Ta4  & 0.66665 & 0.66667 & 0.25000 & 0.00967(28) \\
Ta5  & 0.66646 & 0.33329 & 0.25000 & 0.00839(28) \\
Ta6  & 0.66665 & 0.00004 & 0.25000 & 0.00892(28) \\
\hline
Se1  & 0.00000 & 0.77782 & 0.38226 & 0.01148(23) \\
Se2  & 0.00000 & 0.44445 & 0.38251 & 0.01073(23) \\
Se3  & 0.00000 & 0.11111 & 0.38249 & 0.00980(22) \\
Se4  & 0.33333 & 0.11106 & 0.38230 & 0.01062(22) \\
Se5  & 0.33316 & 0.44434 & 0.38232 & 0.01028(22) \\
Se6  & 0.33322 & 0.77786 & 0.38237 & 0.01084(23) \\
\hline
\end{tabularx}
\begin{tabularx}{\linewidth}{@{}l *{3}{>{\centering\arraybackslash}X}@{}}
\hline
\textbf{3R} & \textbf{a(\AA)} & \textbf{b(\AA)} & \textbf{c(\AA)} \\
\hline
  & 3.43103(16) & 3.43103(16) & 19.12068(7)  \\
\end{tabularx}
\begin{tabularx}{\linewidth}{@{}l *{4}{>{\centering\arraybackslash}X}@{}}
\hline
& $\mathbf{x}$ & $\mathbf{y}$ & $\mathbf{z}$ & $\mathbf{U}_{\mathrm{iso}}$ \\
\hline
Ta1  & 0.00000 & 0.00000 & 0.00000 & 0.01224(63) \\
\hline
Se1  & 0.3333  & 0.6667 & 0.25000 & 0.00478(64) \\
Se2  & 0.3333  & 0.6667 & 0.417 & 0.00311(62) \\
\hline
\end{tabularx}
\end{table}

The correlation between interlayer spacing and transition temperatures (CDW and superconductivity) is a key insight.Through a comparative analysis of their structural, transport, and neutron diffraction data, we demonstrate that the nature of c-axis stacking directly influences interlayer coupling, dimensionality, and the balance between CDW order and superconductivity. The 1T phase, with its AA stacking and octahedral coordination, exhibits strong interlayer orbital overlap, stabilizing a high-temperature commensurate CDW phase but at the same time suppressing superconductivity. In contrast, the 2H phase, with AB stacking and trigonal prismatic coordination, supports both incommensurate and commensurate CDW transitions and weak superconductivity at low temperatures. Most notably, the 3R phase, characterized by ABC stacking and non-centrosymmetric structure, shows enhanced superconductivity alongside CDW order, suggesting a reduced competition between these correlated states. The increase in interlayer spacing from 1T to 3R enhances the 2D character of the electronic structure, reduces orbital overlap, and modifies the Fermi surface topology. These factors are critical in tuning the emergence of superconductivity. Unlike the 1T and 2H phases where CDW tends to suppress superconductivity, the 3R phase shows coexistence, suggesting a reduced competition or even a synergistic relationship. This makes 3R-TaSe$_2$ a model system for studying intertwined orders. These findings emphasize the importance of dimensional and structural control in layered transition metal dichalcogenides and open avenues for engineering quantum phases through stacking and symmetry manipulation.
All three polymorphs exhibit a CDW transition because their quasi-2D layered structure and Fermi surface topology promote electron–phonon–driven localization. In 1T, the CDW onset occurs at very high temperatures, indicating that the Fermi surface is gapped which suppresses superconductivity. The 2H and 3R phases superconduct because their prismatic coordination and weaker CDW states preserve enough electronic states at the Fermi level, allowing pairing. 
Our work clearly demonstrates that c-axis stacking (AA, AB, ABC) directly influences interlayer coupling, which in turn governs the balance between CDW and superconductivity. Such structural control provides a tunable pathway to manipulating correlated electronic states in layered materials. It also highlights the impact of stacking sequences and local coordination geometries on the electronic properties of the TaSe${_2}$ polymorphs, 1T, 2H, and 3R. The increase in interlayer spacing from 1T to 3R enhances the 2D character of the electronic structure. This leads to reduced orbital overlap, modified electronic band structures, and enhanced superconductivity in the 3R phase. Specifically, the 3R phase lacks inversion symmetry, possibly enabling Ising superconductivity and potentially unconventional pairing mechanisms, opening avenues for exploring topological superconductivity and spin-orbit coupled quantum states \cite{xie2025}.

\section*{Acknowledgments}
\label{Acknowledgments}
The authors would like to acknowledge support from the National Science Foundation, Grant No. 2219493. A portion of this research used resources at ANSTO. The data can be obtained from the following repository \cite{cdw_materials_repo}. The authors acknowledge Diane Dickie for assistance provided with the single crystal X-ray measurements, and John Schneeloch for help with the sample synthesis. The single-crystal X-ray diffraction experiments were performed on a diffractometer at the University of Virginia, funded by the NSF-MRI program (Grant CHE-2018870). The data are available at \cite{cdw_materials_repo}.
$\ast$To whom correspondence should be addressed: louca@virginia.edu.

\bibliography{bibliography}

\end{document}